\documentclass[12pt, twocolappendix, appendixfloats, numberedappendix]{emulateapj}

\def\eg{e.g., }

\newcommand{\mpc}{{\rm\,Mpc}}
\newcommand{\kpc}{{\rm\,kpc}}

\newcommand{\beq}{\begin{equation}}
\newcommand{\eeq}{\end{equation}}

\def\hyi{\ion{H}{1}}

\def\mgii{\ion{Mg}{2}}

\def\ciii{\ion{C}{3}]}
\def\civ{\ion{C}{4}}

\def\cai{\ion{Ca}{1}}
\def\caii{\ion{Ca}{2}}
\def\caiii{\ion{Ca}{3}}

\def\ovi{\ion{O}{6}}

\def\lymana{Ly$\alpha$}

\def\ha{H$\alpha$}
\def\hb{H$\beta$}

\newcommand{\kms}{\ensuremath{{\rm km~s}^{-1}}}

\newcommand{\rewcaiione}{\ensuremath{W_0^{\rm K}}}

\newcommand{\MSun}{\ensuremath{\rm M_\odot}}
\newcommand{\caplus}{\ensuremath{\rm{Ca^+}}}

\usepackage{natbib}
\usepackage{rotating}

\begin{document}

\shorttitle{Calcium H \& K absorption by galaxy halos}
\shortauthors{Zhu \& M{\'e}nard}
\title {Calcium H \& K induced by galaxy halos}

\author{
Guangtun Zhu\altaffilmark{1} \& Brice M{\'e}nard\altaffilmark{1,2,3}
} 
\altaffiltext{1}{Department of Physics \& Astronomy, Johns Hopkins University, 3400 N. Charles Street, Baltimore, MD 21218, USA, gz323@pha.jhu.edu}
\altaffiltext{2}{Institute for the Physics and Mathematics of the Universe, Tokyo University, Kashiwa 277-8583, Japan}
\altaffiltext{3}{Alfred P. Sloan Fellow}

\begin{abstract}
We present a measurement of the mean density profile of \caii\ gas around galaxies out to $\sim200\,\kpc$, traced by Fraunhofer's H \& K absorption lines. The measurement is based on cross-correlating the positions of about one million foreground galaxies at $z\sim0.1$ and the flux decrements induced in the spectra of about $10^5$ background quasars from the Sloan Digital Sky Survey. This technique allows us to trace the total amount of \caii\ absorption induced by the circumgalactic medium, including absorbers too weak to be detected in individual spectra. We can statistically measure \caii\ rest equivalent widths down to several m\AA, corresponding to column densities of about $5\times10^{10}\,{\rm cm }^{-2}$. We find that the \caii\ column density distribution follows $N_{\rm Ca\,II}\sim r_\mathrm{p}^{-1.4}$ and the mean \caii\ mass in the halo within $200\,\kpc$ is $\sim5\times10^3~M_\odot$, averaged over the foreground galaxy sample with median mass $\sim10^{10.3}\,\MSun$. This is about an order-of-magnitude larger than the \caii\ mass in the interstellar medium of the Milky Way, suggesting more than $90\%$ of \caii\ in the Universe is in the circum- and inter-galactic environments. Our measurements indicate that the amount of \caii\ in halos is larger for galaxies with higher stellar mass and higher star formation rate. For edge-on galaxies we find \caii\ to be more concentrated along the minor axis, i.e. in the polar direction. This suggests that bipolar outflows induced by star formation must have played a significant role in producing \caii\ in galaxy halos.
\end{abstract}


\section {Introduction}


The H \& K absorption and emission lines have had an important role in the early developments of spectroscopy and, for about two centuries, have been key to understanding a host of astronomical phenomena ranging from the Solar atmosphere to the interstellar medium of distant galaxies.

The labelling convention for the major absorption lines traces its origin back to the work of the German optician and glass manufacturer Joseph Fraunhofer who, in 1814, catalogued over 500 lines in the solar spectrum and designated the principal dark features with letters from A through H \citep{fraunhofer14a}. He used the letter I to denote the blue end of the visible spectrum and labelled a selected set of weaker lines with lower-case letters. Despite being one of the strongest features in the solar spectrum, the bluer line of the doublet was never labelled by Fraunhofer. His main goal was to measure the refractive indices of various materials and was probably not interested in using pairs of lines closely separated in wavelength. The nineteenth-century literature refers to this second line as H2, H' or HII. 
According to \citet{pasachoff10a}, the origin of the modern labelling goes back to John William Draper who in 1843 named the lines as H and k (lower case) in order of decreasing wavelength \citep{draper43a}. The first person to use the capital K to designate the second H line is \'Eleuth\`ere Mascart from the Ecole Sup\'erieure de Paris in 1864 \citep{mascart64a}. The naming convention was straightened out at the first International Astronomical Union meeting, held in Rome in 1922. The Committee accepted that only some of the mostly century-old symbols should be retained. It was decided that only A ($\rm{O_2}$), a ($\rm{O_2}$), D (Na), b (Mg), G (Fe \& Ca), H and K were to be preserved and the notation \ha, and \hb, etc. should be adopted for the hydrogen lines \citep{hearnshaw86a}.

The H \& K lines have played a major role in astrophysics because of their strength and their location in the visible part of the spectrum. They correspond to the fine structure splitting of the singly ionized calcium excited states \caii\ (\caplus) at $\lambda=3934.78\,$\AA\ (K) \& $3969.59\,$\AA\ (H)  in vacuum. Being an abundant element, $\log(\rm{Ca/H})_\odot+12 \simeq 6.34$ \citep{asplund09a}, calcium is found in a range of astrophysical environments. Calcium is a refractory element and has ionization potentials of 6.11 and 11.87 eV, for \cai\ and \caii, respectively \citep{morton03a}. In high-density environments such as the disk of our Galaxy, most of the calcium is found in dust grains. In low-density environments, it tends to be found in \caiii\ and other higher-ionization states.

\begin{figure}[t]
\centering
\epsscale{1.25}
\plotone{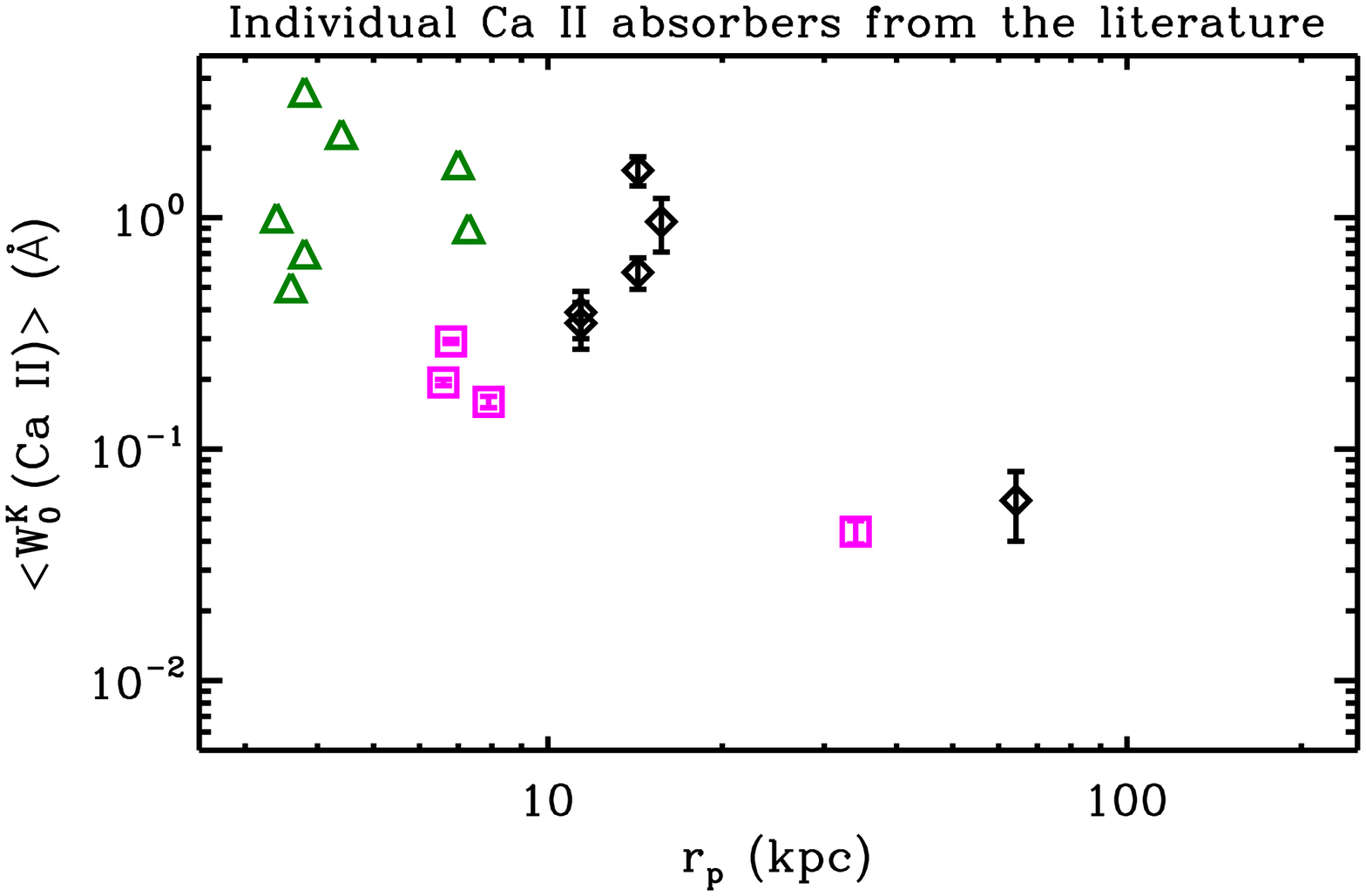}
\caption{Individual \caii\ absorbers with known associated galaxies, compiled from \citet[][green triangles]{york12a}, \citet[][magenta squares]{richter11a}, and \citet[][black diamonds]{bowen91a}. Only those beyond impact parameter $r_\mathrm{p}=3\,\kpc$ are shown, among which two have $r_\mathrm{p}$ larger than $20\,\kpc$.}
\vspace{0.2cm}
\label{fig:literature}
\end{figure}

The H \& K lines are the only resonance lines in cool-star spectra of the dominant ionization stage of an abundant element that are accessible to ground based observations. In 1814, Fraunhofer himself detected these absorption lines in the spectra of various stars. About a century later, \citet{hartmann04a} detected H \& K absorption due to interstellar matter, opening a new window on the mapping of the gas distribution in the Galaxy. On extragalactic scales, \caii\ absorption was detected by \citet{boksenberg78a} in the spectrum of the quasar 3C 232. The first detections of the H \& K absorption due to galaxy halos followed \citep[][]{boksenberg78a, boksenberg80a, blades81a}. Figure~\ref{fig:literature} shows a compilation of the detections reported in the literature. In such low-density environments, the mapping of \caii\ has been limited to the immediate vicinity of galaxies, i.e. within about 20 kpc due to the weakness of these absorption lines. The first detection of calcium H \& K absorption by a high-velocity cloud (HVC) was reported by West et al. (1985), showing the Milky Way halo is enriched with metals. In this paper we extend the use of the H \& K lines to the circum- and inter-galactic context. 

We present a measurement of the mean density profile of \caii\ in galaxy halos out to $\sim200$~kpc. We are also able to explore its dependence on stellar mass, star formation rate (SFR), and azimuthal angle, and place strong constraints on both the total amount of \caii\ in galaxy halos and its origins. Our statistical analysis is based on cross-correlating the positions of about one million foreground galaxies at $z\sim0.1$ and the flux decrements induced in the spectra of about $10^5$ background quasars from the Sloan Digital Sky Survey \citep[SDSS, ][]{york00a}. Metals in galaxy halos can have different origins - enriched outflows driven by star formation or active galactic nuclei, gas stripped out of satellite galaxies by ram pressure or tidal force or outflows, gas in satellite galaxies, and enriched infalling intergalactic gas. Probing the distribution of \caii\ may shed new light on these processes.

The paper is organized as follows: In Section $2$, we briefly describe the dataset and the method. In Section $3$, we present the results of the statistical analysis, estimate the total \caii, calcium and gas mass, and explore their dependence on galaxy properties. Section $4$ summarizes our results. We assume the $\Lambda$CDM cosmogony, with $\Omega_\Lambda=0.7$, $\Omega_{\mathrm m}=0.3$, and ${\mathrm H}_0=70\,\kms\,\mpc^{-1}$.

\section{Data analysis}\label{sec:data}

\begin{figure*}
\epsscale{1.0}
\plotone{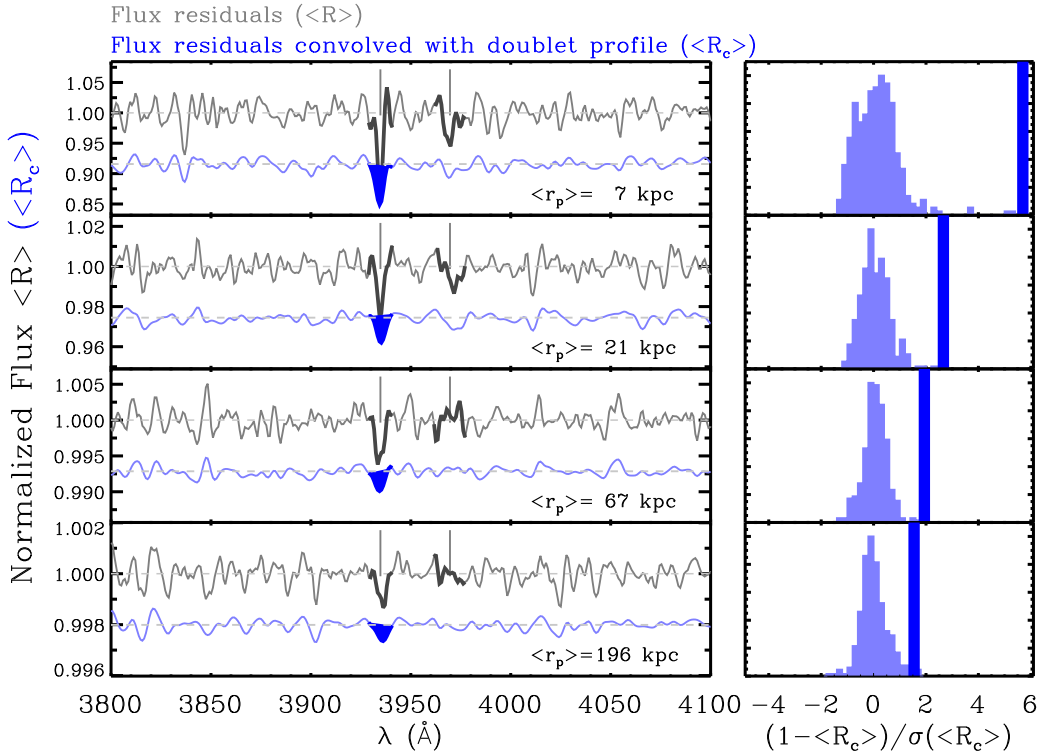}
\caption{Stacked continuum-normalized spectra of background quasars. In the left panel, the gray lines show the geometric means of original residual spectra, while the blue lines show the geometric means of residual spectra convolved with a double Gaussian profile.
In the right panel, we compare the \caii\ signals (dark blue vertical line) with the distributions of the flux residuals (light blue histogram).
}
\label{fig:stackspec}
\end{figure*}

Our goal is to statistically measure the H \& K absorption lines imprinted in the spectra of distant quasars by the circumgalactic medium (CGM) of low-redshift galaxies and estimate the mean column density of \caii\ within their virial radius and beyond. To do so we measure the mean rest equivalent width, $\langle \rewcaiione \rangle$, of the \caii\ K line imprinted in quasar spectra, as a function of impact parameter (projected galactocentric distance, $r_\mathrm{p}$) from low-redshift galaxies. A similar technique has been used at higher redshifts to probe various UV absorption lines by \citet{steidel10a} with the Keck Baryonic Structure Survey and \citet{bordoloi11a} using the zCOSMOS survey \citep[][]{lilly07a}.

When \caii\ absorption is in the linear regime of the curve of growth, the rest equivalent width is related to \caii\ column density with 
\beq
N_{\rm Ca\,II} 
= 
1.13 \times 10^{20}~\mathrm{cm}^{-2}~\frac{\rewcaiione}{f\lambda^2}\, \mathrm{,}
\label{eq:density}
\eeq
where both $\rewcaiione$ and $\lambda$ are in unit of \AA\ and the oscillator strength $f$ for the \caii\ K line is $0.648$ \citep[][]{safronova11a}. This relation is valid when the optical depth for \caii\ K absorption is smaller than unity, which corresponds to {\bf $N_{\rm Ca\,II}\lesssim 10^{12.5}\,{\rm cm}^{-2}$} (assuming the Doppler broadening factor $b\sim10\,\kms$)\footnote{From high-resolution spectroscopy of individual \caii\ absorbers, the median Doppler broadening factor $b\sim6\,\kms$ \citep[\eg][]{richter11a}. Usually there are several unsaturated components induced by the CGM of a single galaxy, and we cannot resolve these different components here because SDSS has an instrumental resolution $\sim 69\,\kms$. The \emph{apparent} Doppler broadening factor therefore is dominated by the separation of different components, which is $\gg 10\,\kms$.}. As shown below, this is the regime of interest in the present study.

Our analysis makes use of data from SDSS DR7 \citep[][]{abazajian09a}. We use the value-added MPA-JHU catalog \citep[][]{kauffmann03a, brinchmann04a}\footnote{\tt http://www.mpa-garching.mpg.de/SDSS/DR7/}, which includes stellar mass and SFR measurements for $819,437$ unique galaxies with a mean redshift $z\sim0.1$. We also supplement the catalog with position angle and minor-to-major axis ratio ($b/a$), retrieved from SDSS skyserver\footnote{\tt http://cas.sdss.org/astrodr7/en/}. To ensure that the induced \caii\ absorption is not blended with that of our Galaxy, we select galaxies at redshifts $z_{\rm gal}>0.02$. We select quasars as background sources \citep{schneider10a} and use the improved redshift estimates given by \citet[][]{hewett10a}\footnote{\tt http://das.sdss.org/va/Hewett\_Wild\_dr7qso\_newz/}. The sample includes $107,194$ quasars with redshifts spanning $0.1<z<6.5$. 

Absorption measurements are based on continuum-normalized fluxes:
\beq
R(\lambda) = \frac{F(\lambda)}{\hat F_{\rm cont}(\lambda)}\, \mathrm{.}
\label{eq:residual}
\eeq
For the extraction of weak absorption information from the noise-dominated flux residuals, accurate estimation of the source flux continuum is crucial. We make use of the SDSS quasar continuum estimates ${\hat F_{\rm cont}(\lambda)}$ given by \citet[][]{zhu13a}. In a nutshell, this analysis employed the robust dimensionality-reduction technique {\it nonnegative matrix factorization} \citep[NMF,][]{lee99a, blanton07a} to construct a basis set of nonnegative quasar eigenspectra, and fit each observed quasar spectrum with a nonnegative linear combination of these eigenspectra. Large-scale residuals not accounted for by the NMF basis set were removed with appropriate median filters. For details regarding this continuum estimation method we refer the reader to \citet{zhu13a}. The continuum estimation requires robust NMF basis, which restricts the quasar sample to those at $z<4.7$.  In total we are able to use $84,533$ quasar spectra. Finally, when selecting background quasars, we require $z_{\rm quasar}-z_{\rm gal}>0.1$.

As we are intersted in estimating absorption equivalent widths in the rest-frame of selected galaxies at various redshifts, we will shift every flux residual $R(\lambda)$ to the rest frame of the absorbing material throughout the paper. To optimally extract information on absorption lines we use a matched-filter technique \citep[\eg][]{schneider93a} to convolve the residuals $R(\lambda)$ with the expected absorption line profile $P(\lambda)$:
\beq
R_c(\lambda) = 1-\frac{\int [1-R(\lambda')] P(\lambda'-\lambda)\,\mathrm{d}\lambda'}{\int P^2(\lambda'-\lambda)\,\mathrm{d}\lambda'}\ \mathrm{,}
\label{eq:convolution}
\eeq
where $P(\lambda)$ is normalized with
\beq
\int P(\lambda)\,\mathrm{d}\lambda = 1\, \mathrm{.}
\label{eq:normalization}
\eeq
When focusing on a single absorption line, $P(\lambda)$ is chosen to be a Gaussian kernel, with a fixed velocity dispersion of $3\,{\rm pixels}$ ($\sim207\,\kms$). We note the exact value of the velocity dispersion adopted ($2$, $3$, or $4$ pixels) has little effect on the results. To extract information from the \caii\ doublet, $P(\lambda)$ is a double-Gaussian kernel with a $34.8$\,\AA\ separation and a doublet ratio of 2 and for which $\lambda$ is the central wavelength of the blue K line. The convolved flux residuals can be used to obtain a non-parametric measurement of the \caii\ equivalent width. The K-line rest equivalent width is given by
\beq
\rewcaiione = 
\frac{2}{3}\,\left[ 1-R_c(\lambda_\mathrm{K}) \right]\, \mathrm{.}
\label{eq:R_to_W_obs}
\eeq

When measuring the mean \caii\ rest equivalent width induced by a sample of foreground galaxies, we select quasars for which the expected \caii\ absorption feature does not fall in the \lymana\ forest. We also exclude quasars for which the expected \caii\ absorption overlaps the quasar's \civ, \ciii, and \mgii\ emission-line regions, although including these systems in the analysis has little effect on our results.

\section{Results}\label{sec:results}

\begin{figure*}
\epsscale{1.0}
\plotone{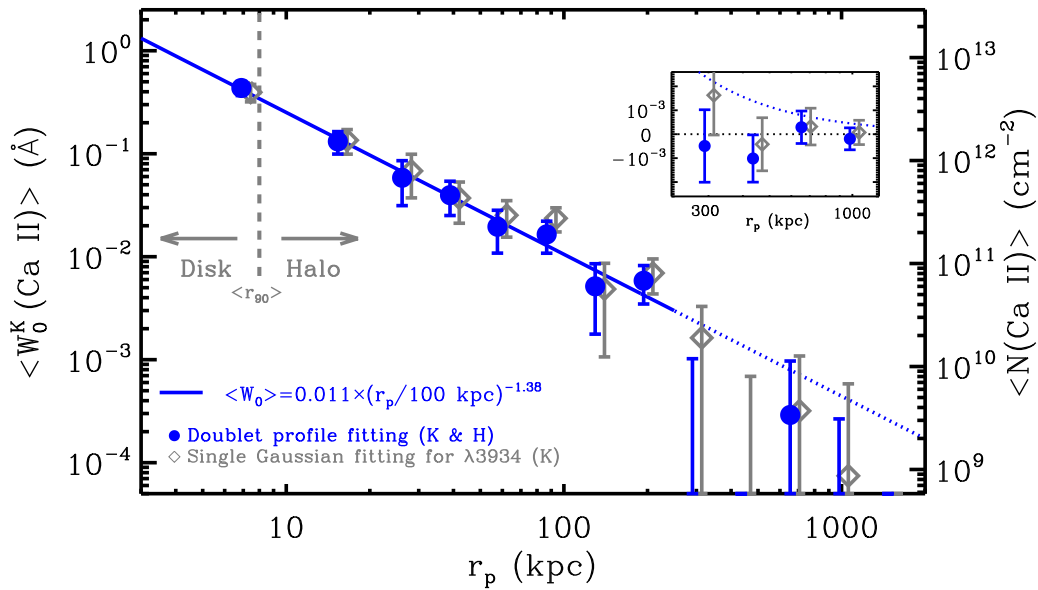}
\caption{The rest equivalent width of \caii\ as a function of impact parameter. The blue solid circles show the measurements using spectra convolved with the double Gaussian kernel. The gray open diamonds are measurements using K line only (with a single Gaussian kernel). The solid line shows the best-fit power law to the measurements at $r_\mathrm{p}<200\,\kpc$, while the dotted line is the extrapolation of the best-fit profile. The inset shows the measurements and the extrapolation of the best-fit profile on large scales, but in linear scale in $y$-axis. The vertical dashed line shows the mean $90\%$ light radius of the foreground galaxy sample ($\langle r_{90} \rangle \sim8\,\kpc$).
}
\label{fig:profile}
\end{figure*}

\subsection{The average \caii~absorption radial profile}\label{sec:profile}

Our estimate of the mean \caii\ absorption is given by the cross-correlation between the position of foreground galaxies denoted by $\delta_\mathrm{g}(r)$ and the absorption rest equivalent width expected in the spectra of quasars, as a function of projected separation $r_\mathrm{p}$:
\beq
\langle \rewcaiione \rangle  (r_\mathrm{p})
=
\left \langle
\delta_\mathrm{g} \;.\; \rewcaiione(r_\mathrm{p})
\right \rangle \, \mathrm{,}
\label{eq:main}
\eeq
where the ensemble average is taken at the position of every galaxy and includes all background quasars in a radial bin centered on $r_\mathrm{p}$. Absorption being a multiplicative effect we use a geometric mean for the ensemble average, which gives us an arithmetic mean of the corresponding optical depth. However, using an arithmetic mean yields similar results, as expected when measuring weak absorption lines. Our estimator is inverse-variance weighted, using the wavelength-dependent noise given by the SDSS pipeline.

In Figure~\ref{fig:stackspec} we show intermediate measurements leading to our estimate of $\langle \rewcaiione \rangle (r_\mathrm{p})$. 
The left panel shows, for four radial bins, the stacked galaxy rest-frame \& continuum-normalized quasar fluxes $\langle R(\lambda) \rangle$ between $3800$ and $4100$\,\AA\ (gray lines). The vertical lines show the expected locations of the H \& K lines. The blue lines show the convolved residuals $\langle R_\mathrm{c}(\lambda) \rangle$. To estimate the uncertainty, we measure the distribution of the convolved residuals $\langle R_\mathrm{c}(\lambda) \rangle$ over the wavelength range between $3800$ and $4100$\,\AA\ and estimate its dispersion. This is illustrated in the right panel of the figure, where the vertical lines indicate the corresponding values of $\langle R_\mathrm{c}(\lambda_\mathrm{K}) \rangle$. 

\begin{deluxetable}{cccc}
\tabletypesize{\scriptsize}
\tablecolumns{4}
\tablecaption{The average Ca II absorption radial profile}
\tablehead{
 \colhead{$r_\mathrm{p}$ bin} &  \colhead{Median $r_\mathrm{p}$} &  \colhead{$N_{\rm pairs}$} & \colhead{$\langle W_0^{\mathrm{K}}({\rm Ca\,II}) \rangle$} \\
 \colhead{[kpc]} &  \colhead{[kpc]} & \colhead{} & \colhead{[${\rm m\AA}$]}\\
}
\startdata
$(3, 10]$   & $7$   & $17$    & $435\pm68$    \\
$(10,20]$   & $15$  & $86$    & $132\pm33$    \\
$(20,30]$   & $26$  & $131$   & $59\pm27$     \\
$(30,45]$   & $39$  & $410$   & $40\pm15$     \\
$(45,68]$   & $58$  & $1106$  & $19.5\pm8.7$  \\
$(68,101]$  & $87$  & $2991$  & $16.5\pm5.7$  \\
$(101,152]$ & $130$ & $7456$  & $5.2\pm3.4$   \\
$(152,228]$ & $194$ & $17348$ & $5.8\pm2.4$   

\enddata
\label{tbl:npairs}
\end{deluxetable}


\begin{figure*}[!t]
\epsscale{0.57}
\plotone{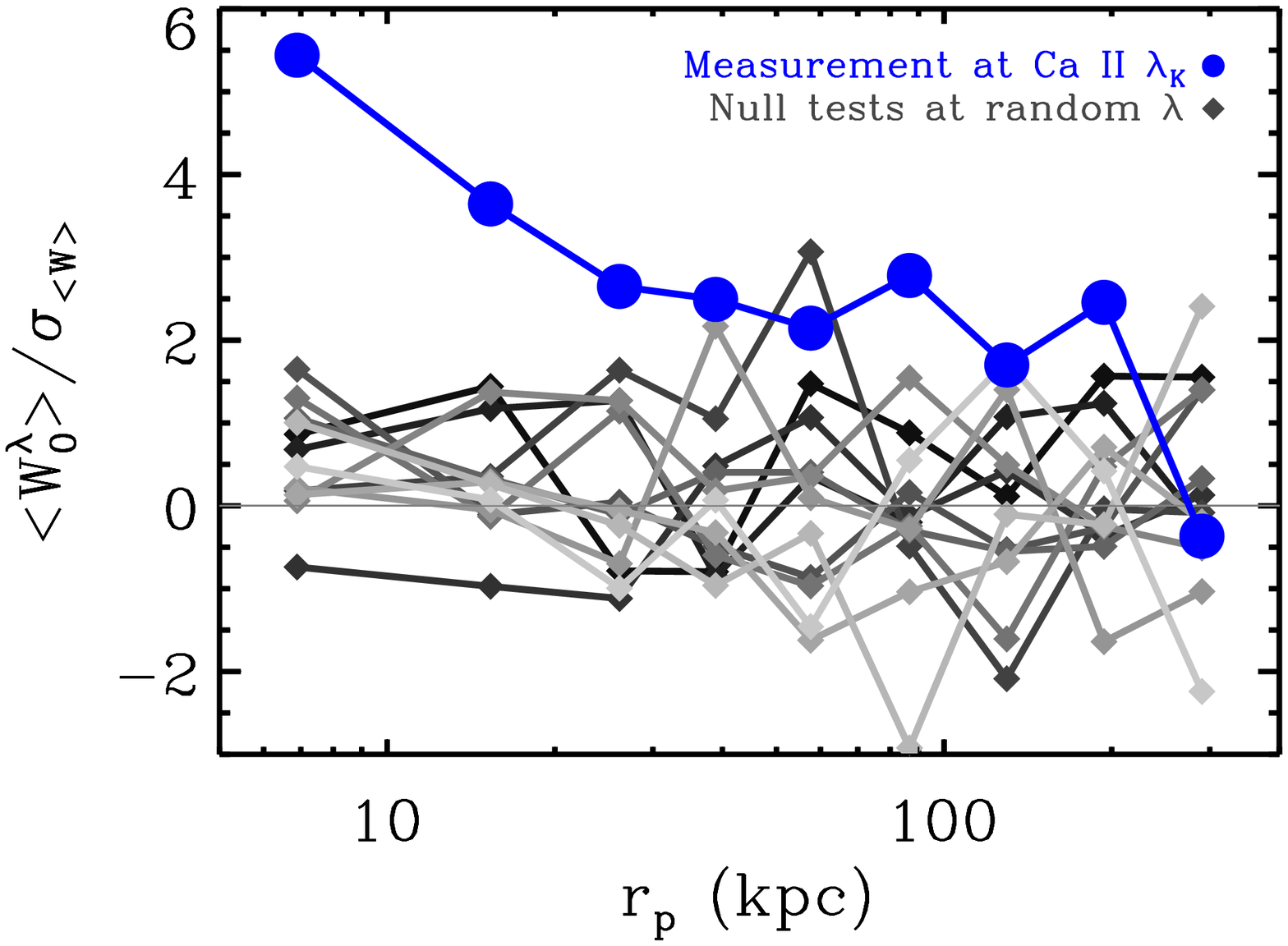}
\epsscale{0.57}
\plotone{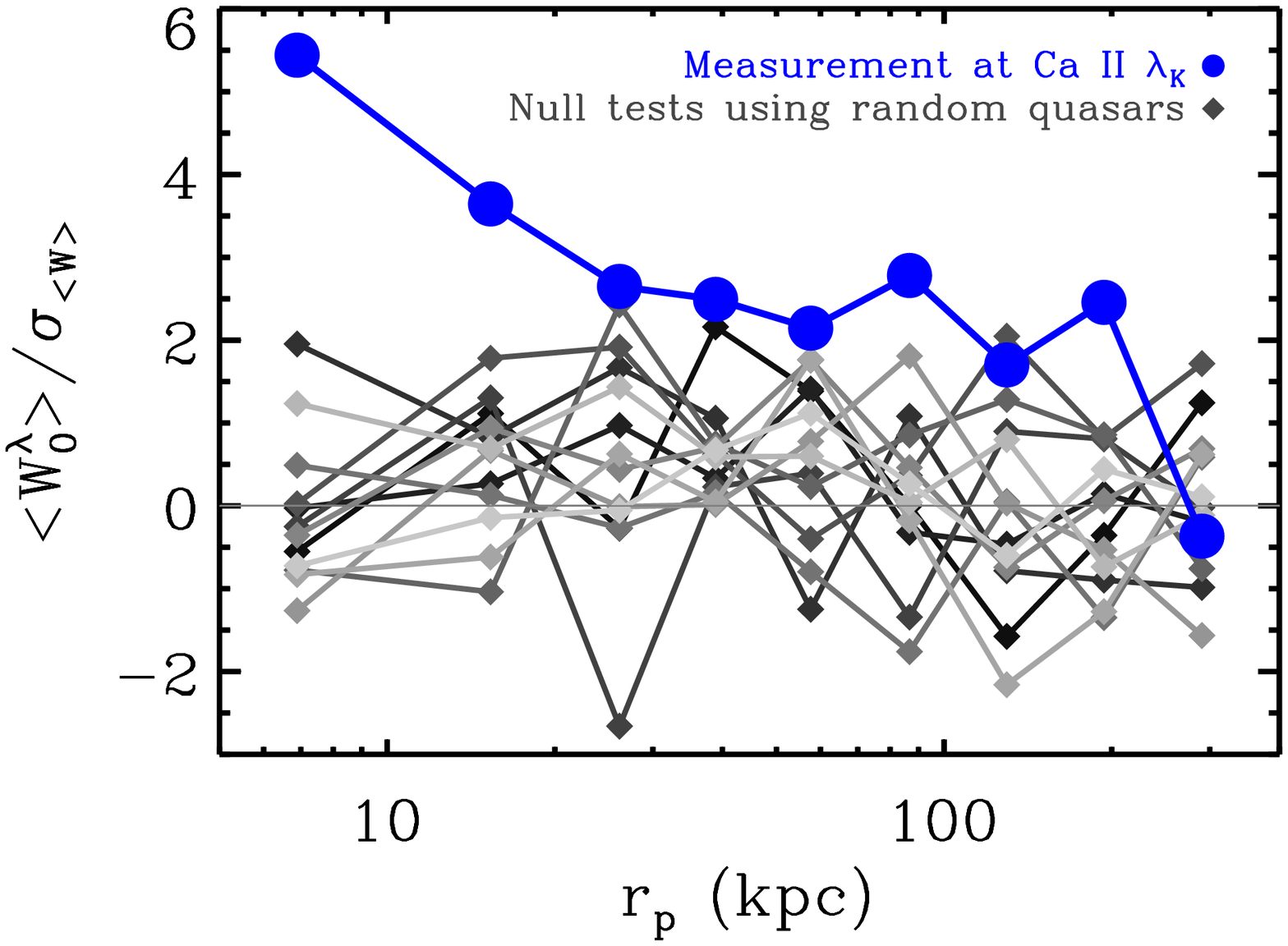}
\caption{Null hypothesis tests showing the robustness of the \caii\ detection. {\it Left} panel: significance of rest equivalent width measurements at randomly-selected wavelengths. {\it Right} panel: significance of rest equivalent width measurements but using random quasars at similar redshifts as those in galaxy-quasar pairs. The blue solid circles show the significance of the \caii\ absorption measurements.
}
\label{fig:nulltest}
\end{figure*}

We present the radial dependence of \caii\ absorption around galaxies in Table~\ref{tbl:npairs} and Figure~\ref{fig:profile}. The blue solid circles represent the rest equivalent width estimates $\langle \rewcaiione \rangle$  derived from measurements of the H \& K lines $\langle R_\mathrm{c}(\lambda_\mathrm{K}) \rangle$. We also show the estimates based on the residuals of the K line only ($\langle R(\lambda_\mathrm{K}) \rangle$, with a single Gaussian kernel) with the gray open diamonds. The conversion to column density is shown in the right $y$-axis of the figure. It is important to note that even if some overlap might exist between the angular apertures used around different galaxies, the measurements on different scales are largely uncorrelated as different galaxies tend to be at different redshifts. The inset of the figure shows, using a linear scale, the measurements on scales greater than $200~\kpc$. Our results show that, on average, \caii\ gas is present in the halo of galaxies up to scales comparable to their virial radius. For reference, the vertical dashed line shows the mean value of $r_{90}$, the radius encompassing 90\% of the light of a galaxy, which we take as a fiducial limit between the disk and halo environments. For our sample $\langle r_{90}\rangle \sim8\;$kpc. The median stellar mass of the foreground galaxy sample is $\simeq10^{10.3}\,\MSun$.

To validate the robustness of our measurements, we perform two null hypothesis tests: (1) Using Eq.~\ref{eq:main} we measure the rest equivalent widths at different wavelengths. The left panel of Figure \ref{fig:nulltest} shows the significance of the measurements for $12$ randomly-selected wavelengths between $3800$ and $4100$\,\AA\ (gray diamonds). For comparison, the \caii\ measurements are shown as blue solid circles.
(2) We measure the expected \caii\ rest equivalent widths but, instead of using the corresponding background quasars lying within some impact parameter from the galaxies, we use random quasars with similar redshifts. These results are shown in the right panel of the Figure. In both cases our test measurements are consistent with no signal and show that the detection of \caii\ absorption above (Table~\ref{tbl:npairs} and Figure~\ref{fig:profile}) is robust and not induced by systematic effects.

\subsection{Total mass of \caii\ in the halo}
\label{sec:totalmass}

We parametrize the radial dependence of the mean \caii\ absorption around galaxies with a power-law functional form:
\beq
\langle \rewcaiione \rangle (r_\mathrm{p}) = {\rm A}\,\bigg( \frac{r_{\mathrm p}}{100~\kpc} \bigg)^{\alpha}\, \mathrm{,}
\label{eq:profile}
\eeq
where $r_\mathrm{p}$ is taken as the median impact parameter in each bin. Performing a least-squares fit to the measurements at $r_\mathrm{p}<200\,\kpc$, we obtain:
\beq
{\rm A}=11\pm2\,\text{m\AA}\ \mathrm{and} \ \alpha=-1.38\pm0.11\, \mathrm{.}
\label{eq:profile_fit}
\eeq
We show the best-fit power law with the blue line in Figure~\ref{fig:profile}, where we also show its extrapolation to larger scales. The slope shows the mean \caii\ distribution is steeper than an isothermal profile. Converting $\langle  \rewcaiione \rangle$ to mean column density $\langle N_{\rm Ca\,II} \rangle$ using Eq.~\ref{eq:density} and integrating the best-fit profile between $10$ and $200\,\kpc$, we estimate the average total \caii\ mass in the halo to be \footnote{We note this {\it cylindrical} mass is obtained using projected column density. For spherical mass we need to deproject it through inverse Abel's integral with uncertain assumptions of gas distribution on large scales. The correction factor is $\sim0.6$ and the average spherical mass is $\sim3\times10^3\,\MSun$ assuming the power-law distribution extends to infinity. Throughout the paper we will stick to the observable cylindrical mass.}:
\begin{eqnarray}
\langle M_{\rm Ca\,II}^{\mathrm{halo}} \rangle
&=& 2\,\pi\;\int_{10\,\kpc}^{200\,\kpc} \langle N_{\rm Ca\,II} \rangle (r_\mathrm{p})\;r_\mathrm{p}\,\mathrm{d}r_\mathrm{p}
\nonumber\\
&=& (5.0\pm1.0) \times 10^3\,\MSun\, \mathrm{.}
\label{eq:totalmass}
\end{eqnarray}

It is interesting to compare the total \caii\ mass in the halo with the total \caii\ mass in galaxies themselves such as the Milky Way. To estimate the total \caii\ mass in the Milky Way disk, we use the average neutral hydrogen \hyi\ surface density $10\,\MSun\,\mathrm{pc}^{-2}$ \citep[$\sim10^{21}\,\mathrm{cm}^{-2}$, \eg][]{kalberla09a}, and the typical \caii-to-\hyi\ ratio $N_{\rm Ca\,II}/N_{\rm H\,I}\sim10^{-8.9}$ in the interstellar medium \citep[\eg][]{gudennavar12a}. Integrating the surface density out to $15\,\kpc$, we obtain $M_{\rm Ca\,II}^\mathrm{MW\,Disk} \sim 3\times10^2\,\MSun$, an order-of-magnitude smaller than the average total \caii\ mass in the halo, suggesting more than $90\%$ of \caii\ in the Universe is distributed in circum- and inter-galactic environments. This dramatic difference, however, is likely caused by the difference in the dust depletion levels rather than in the total amount of calcium in different environments. From observations of the Milky Way halo, the gas density in the CGM is more than an order-of-magnitude lower than in the interstellar environment, therefore there is relatively less calcium depleted onto dust grains \citep[\eg][]{wakker00a}, resulting in a \caii\ abundance an order-of-magnitude higher. Finally, we note that the total \caii\ mass in the halo above is obtained by averaging over all the foreground galaxies (in the galaxy-quasar pairs), whose median mass is $\sim 10^{10.3}\,\MSun$.

\subsection{\hyi\ mass traced by \caii}
\label{sec:HI}

We now attempt to estimate the total amount of \hyi\ gas traced by \caii.
It is known that in cool, dense clouds most of Ca is depleted onto grains and \caii\ can be the dominant ionization state, while in warm, lower-density medium, Ca is less depleted but the dominant ionization state is usually \caiii\ \citep[\eg][]{routly52a, welty96a, wakker00a, benbekhti12a}. Here we make the assumption that the \caii-to-\hyi\ in the halo is similar to those observed for high-velocity clouds/Magellanic Stream in the Milky Way halo \citep[$N_{\rm Ca\, II}/N_{\rm H\,I}\sim10^{-7.2}$,][]{wakker01a}. In this case the total amount of \hyi\ traced by \caii\ in galaxy halos is about ${\rm several}\;\times10^9\,\MSun$. For comparison, the total amount of \hyi\ in the Milky Way halo, combining the Magellanic system (Magellanic Stream, the Leading Arm, Magellanic Bridge, excluding galaxies themselves, $3\times10^8\,\MSun$), LMC/SMC interstellar medium ($7\times10^8\,\MSun$), and HVCs ($3\times10^7\,\MSun$), is estimated to be $\sim10^9\,\MSun$ \citep[][]{putman12a}. The total \hyi\ mass in M31's halo, without counting gas in M33, is $\sim\,{\rm several}\times 10^7\,\MSun$ \citep[\eg][]{thilker04a}.

\begin{figure}
\epsscale{1.35}
\plotone{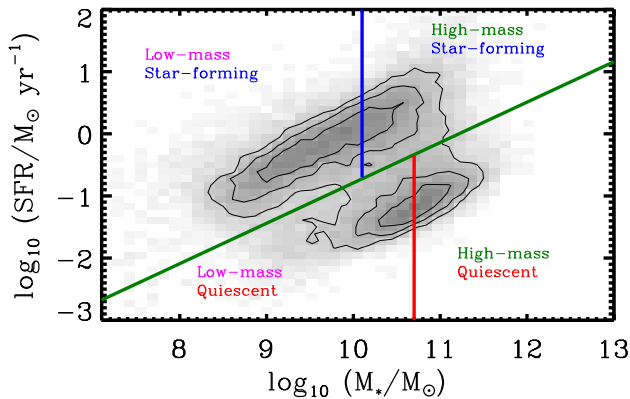}
\caption{Stellar mass-SFR diagram of foreground galaxies in galaxy-quasar pairs. The green line separates star-forming and quiescent galaxies. The blue vertical line ($M_* = 10^{10.1}\,\MSun$) then divides star-forming galaxies into high-mass and low-mass star-forming subsamples, and the red vertical line ($M_*=10^{10.7}\,\MSun$) divides quiescent galaxies into high-mass and low-mass quiescent subsamples.}
\label{fig:masssfrdiagram}
\end{figure}

\begin{figure*}
\epsscale{0.57}
\plotone{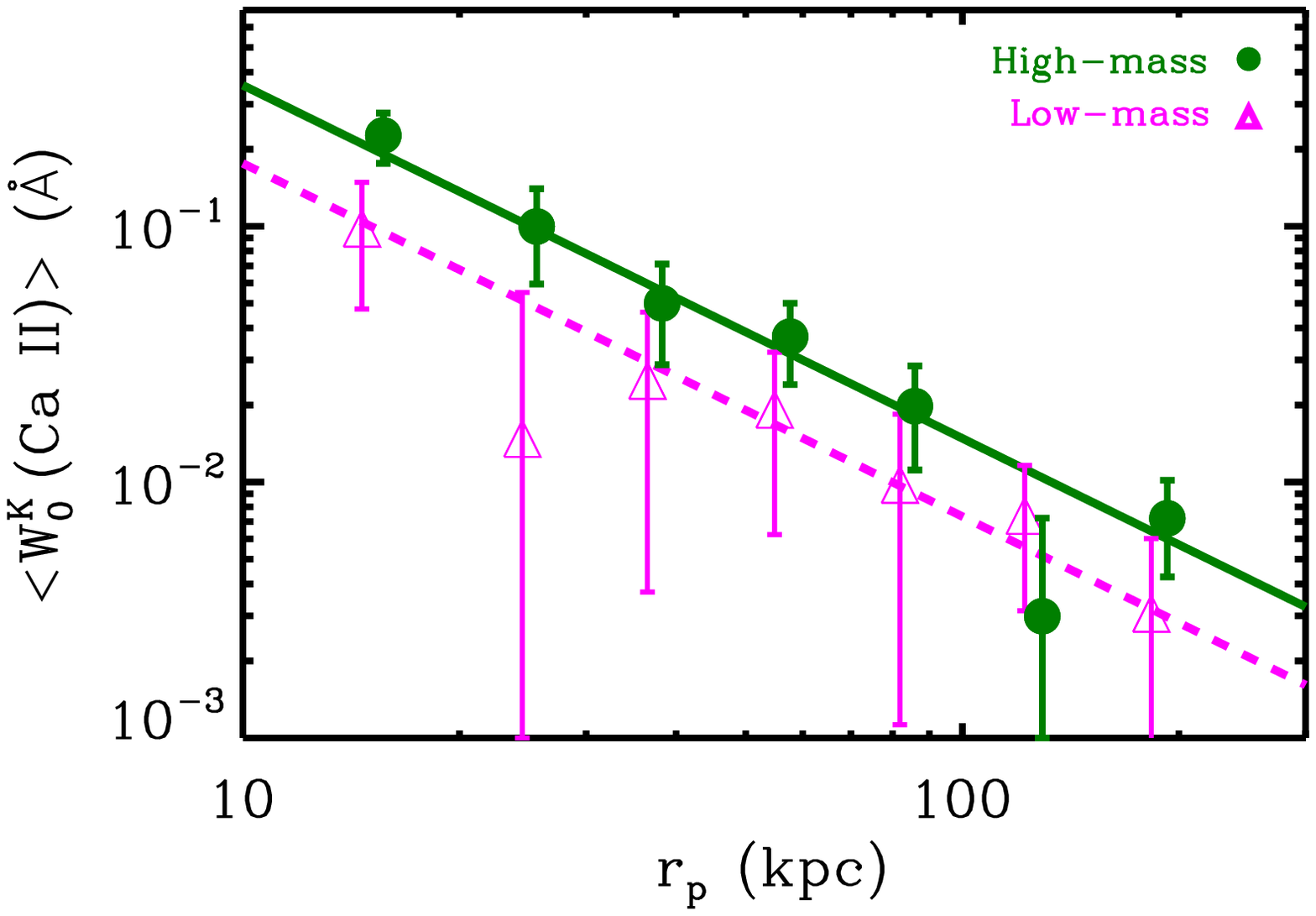}
\epsscale{0.57}
\plotone{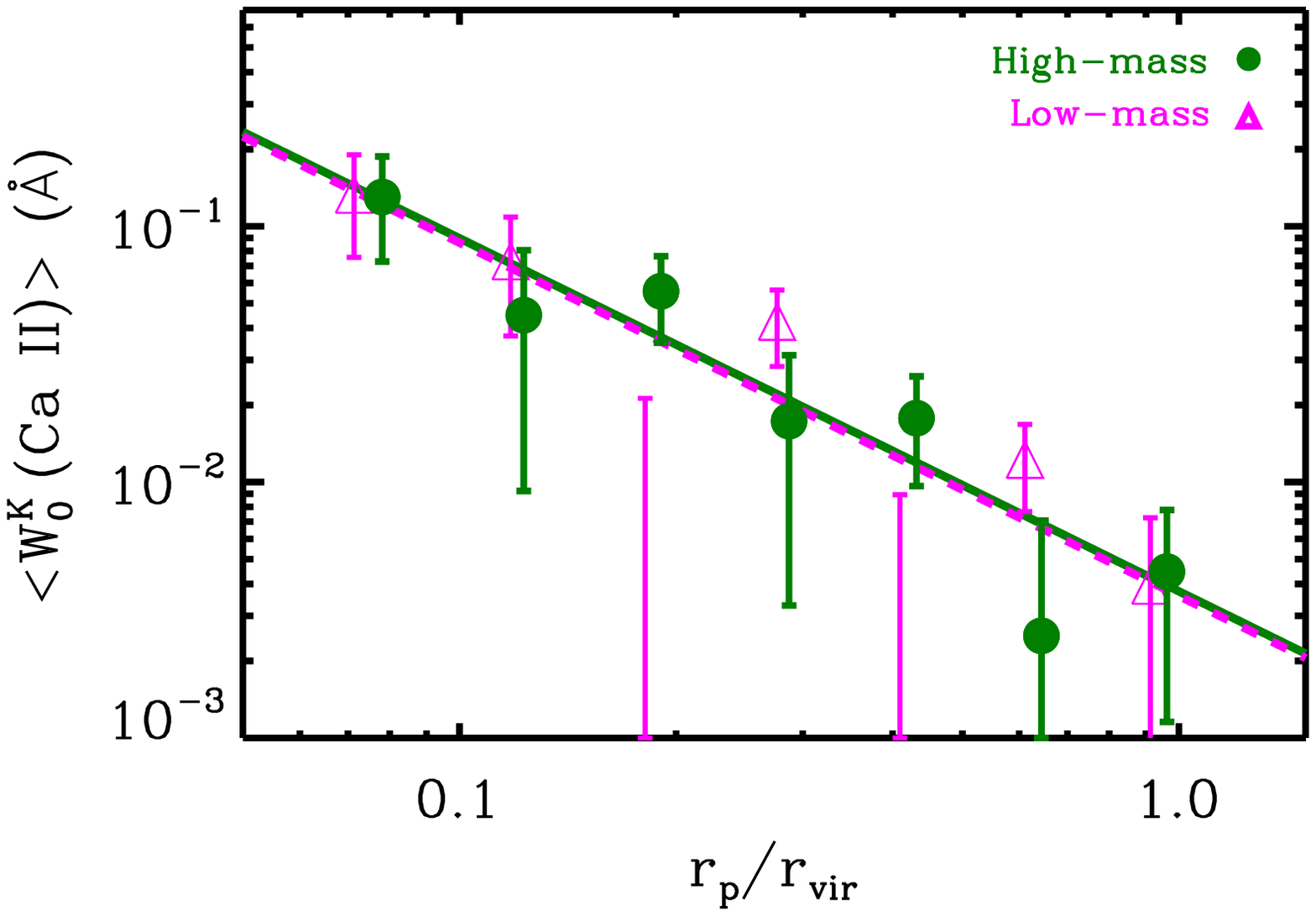}
\caption{Dependence of \caii\ absorption on stellar mass. 
The left panel shows that the \caii\ absorption is stronger around high-mass galaxies (green solid circles) than low-mass ones (magenta triangles). The right panel shows the \caii\ absorption at fixed virial radius-normalized galactocentric distance. The virial radius-scaled \caii\ absorption strengths at different stellar masses appear to be consistent with each other within errors. The corresponding lines show the best-fit power-law profiles with a fixed slope of $\alpha=-1.38$.}
\label{fig:massdepend}
\end{figure*}

\subsection{Dependence on galaxy properties}
\label{sec:dependence}

We now investigate the relation between galaxy properties (stellar mass $M_*$, SFR, and azimuthal angle $\phi$) and the amount of \caii\ in their halos. Considering the bimodal nature of the distribution of galaxies in the stellar mass and SFR space, we divide the sample of foreground galaxies using the following demarcation relation \citep[][]{moustakas13a}: 
\beq
\log_{10}\,{\rm SFR} = -0.79 + 0.65\,(\log_{10}\,M_*-10.0)\, \mathrm{,}
\label{eq:masssfr}
\eeq
where SFR is in $\MSun\,\mathrm{yr}^{-1}$ and stellar mass $M_*$ is in \MSun. This demarcation relation is shown as the green line in Figure \ref{fig:masssfrdiagram} and allows us to separate star-forming galaxies from quiescent ones. To divide each population into high-mass and low-mass subsamples, we adopt a stellar mass separation of $M_*=10^{10.1}$ and $10^{10.7}\,\MSun$ for star-forming and quiescent galaxies as shown with the vertical lines in Figure \ref{fig:masssfrdiagram}. This selection takes into account the fact that star-forming galaxies are on average less massive than quiescent galaxies. We now explore how the amount of \caii\ in the halo depends on the following parameters:\\

\begin{figure*}
\epsscale{0.57}
\plotone{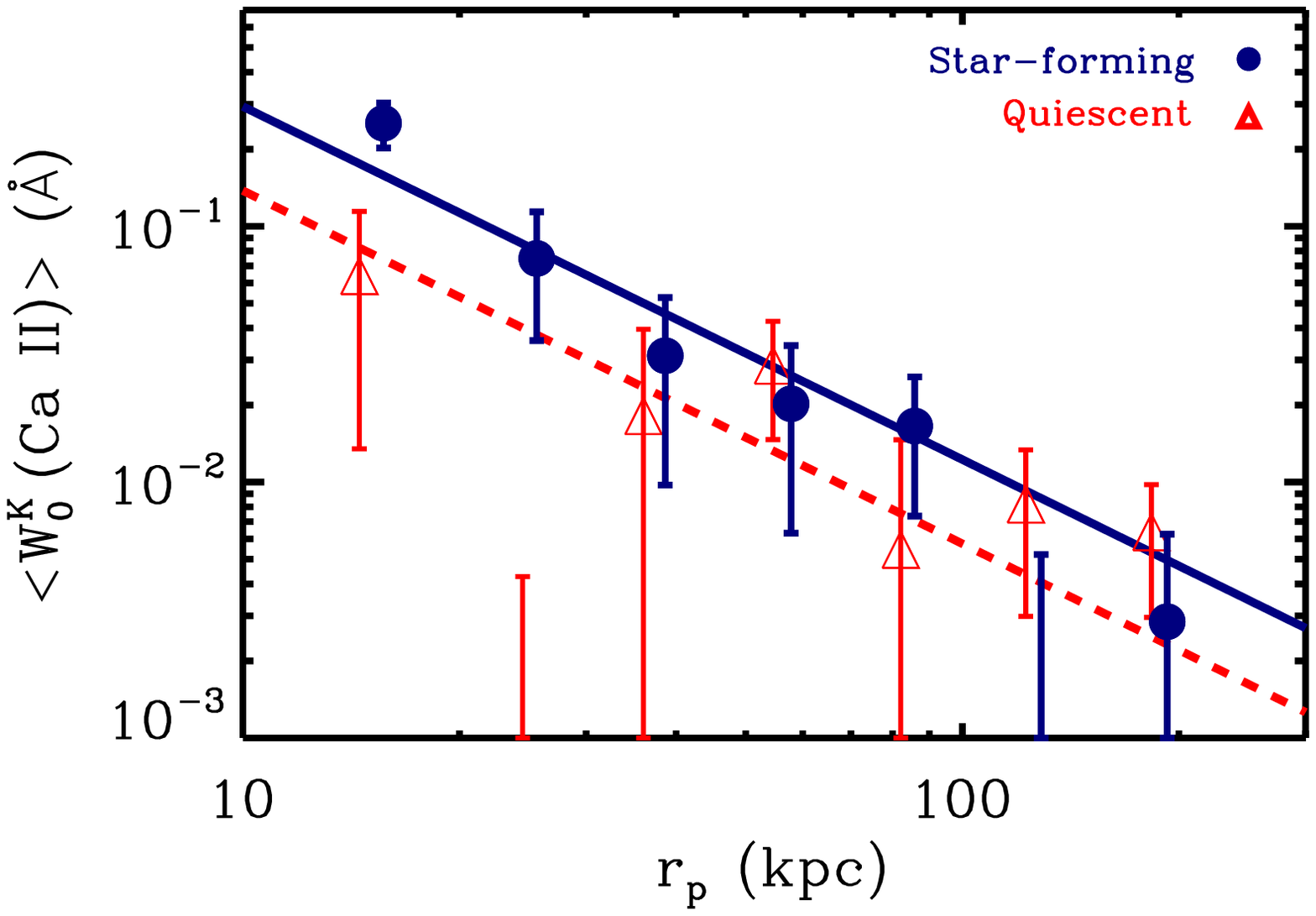}
\epsscale{0.57}
\plotone{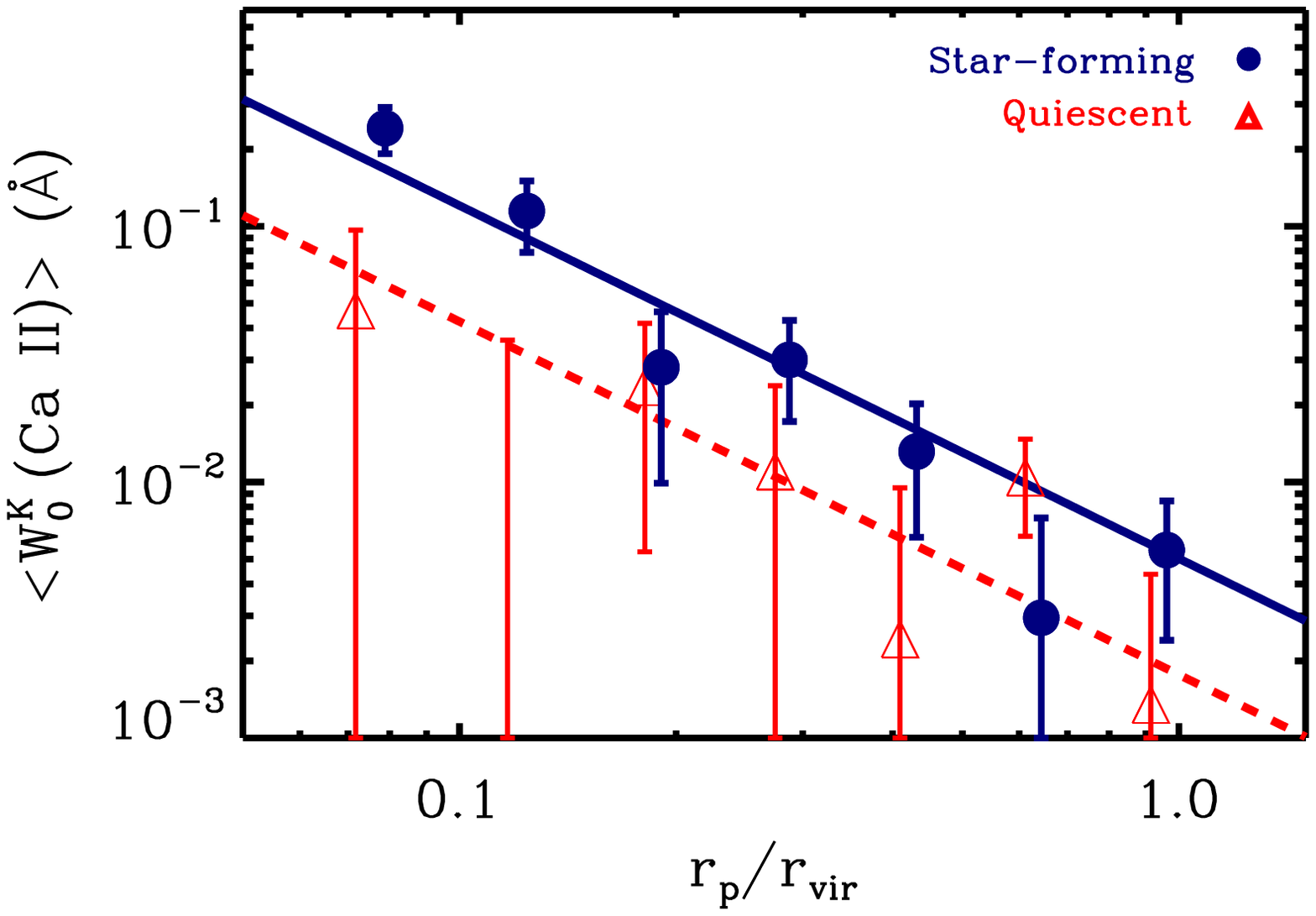}
\caption{Same as Figure~\ref{fig:massdepend}, but for SFR dependence. There is more \caii\ around star-forming galaxies than quiescent ones. The dependence is stronger after normalizing impact parameter with virial radius thus removing the stellar mass dependence.
}
\label{fig:sfrdepend}
\end{figure*}
%

$\bullet~$ {\bf Stellar Mass:} 
we select two samples of galaxies using the mass cuts defined above, combining star-forming and quiescent galaxies. Our measurements show that more massive galaxies have more \caii\ in their halos. To evaluate the difference quantitatively, we fit the \caii\ density profile with a power-law function with a fixed slope $\alpha=-1.38$ as found in Eq.~\ref{eq:profile_fit} (the reduction in S/N prevents us from placing robust constraints simultaneously on both the amplitude and the slope of the measurements). The best-fit amplitudes are
\begin{eqnarray}
A_{\rm high-mass} = 15\pm2\,\text{m\AA} & 
\nonumber\\
A_{\rm low-mass} = 7\pm2\,\text{m\AA} & \mathrm{.}
\end{eqnarray}
For reference, the median stellar masses of high-mass and low-mass populations are $10^{10.5}$ and $10^{9.8}\,\MSun$. Thus, within a fixed impact parameter, the dependence of the amount of \caii\ on stellar mass is shallower than linear.\\

Since the halo of higher-mass galaxies is more extended than that of lower-mass ones, we also look at the dependence on stellar mass by measuring the absorption as a function of impact parameter in unit of virial radius, $r_\mathrm{p}/r_\mathrm{vir}$. To estimate the virial radius $r_\mathrm{vir}$ of a galaxy with a given stellar mass, we assume a power-law relation between virial mass $M_\mathrm{vir}$ and stellar mass $M_*$: 
\beq
M_* \propto M_\mathrm{vir}^{5/3}\;,
\label{eq:virialmass}
\eeq 
where the slope value is taken to be approximately the average slope determined from the {\it abundance matching} technique for galaxies selected between $M_*=10^9$ and $10^{11}\,\MSun$ \citep[\eg][]{behroozi10a}. At a given stellar mass below $10^{11}\,\MSun$, the host dark matter halo mass is statistically independent of morphology (thus SFR), as shown by galaxy-galaxy lensing studies \citep[\eg][]{mandelbaum06a}. The virial radius is then calculated via:
\beq
r_\mathrm{vir}^3 = \frac{3}{4\pi}\frac{M_\mathrm{vir}}{\rho_\mathrm{c}\Delta_\mathrm{c}(z)}  \ \mathrm{,}
\eeq
where $\Delta_\mathrm{c}(z) = (18\pi^2+82x-39x^2)$ and $x=\Omega_\mathrm{m}(z)-1$ \citep[\eg][]{bryan98a}. At the median mass of the $z\sim0.1$ foreground galaxy sample $M_*=10^{10.3}\,\MSun$, $M_\mathrm{vir}\simeq 10^{11.8}\,\MSun$ and $r_\mathrm{vir}\simeq 200\,\kpc$, which is the outer bound we used to calculate the average total mass of \caii\ in the halo for all galaxies. The virial radius is then estimated to be
\beq
r_\mathrm{vir} = 200\,\kpc \ \bigg( \frac{M_*}{10^{10.3}\,\MSun} \bigg)^{1/5}\ \mathrm{.}
\label{eq:rvir}
\eeq
The virial radius-scaled profiles are shown in the right panel of Figure \ref{fig:massdepend}. 
Fitting the dependences with a power-law, keeping the slope $\alpha$, we find the best-fit amplitudes (at $r=0.5\,r_\mathrm{vir}$) are:
\begin{eqnarray}
A'_{\rm high-mass} = 10\pm2\,\text{m\AA} & 
\nonumber\\
A'_{\rm low-mass} = 9\pm2\,\text{m\AA} & \mathrm{,}
\end{eqnarray}
consistent with each other. This shows that the amount of \caii\ in the halo is, on average, more closely related to the dark matter mass than the stellar mass. In other words, the results indicate that the total \caii\ mass within virial radius follows roughly:
\beq
M_{\rm Ca\,II}^{\rm halo} \sim M_\mathrm{vir} \sim M_*^{3/5} \ \mathrm{.}
\label{eq:selfsimilarity}
\eeq
~\\

$\bullet~$ {\bf Star Formation Rate}: we now divide the galaxy sample into star forming and quiescent galaxies using the demarcation represented by the green line in Figure~\ref{fig:masssfrdiagram} (Eq.~\ref{eq:masssfr}). We show the corresponding column density profiles of \caii\ in the left panel of Figure \ref{fig:sfrdepend}. Our measurements show that, on average, star forming galaxies are surrounded by more \caii\ than quiescent systems. We fit the \caii\ density profile with a power-law function fixing the slope $\alpha$ and find the amplitudes to be:
\begin{eqnarray}
A_{\rm star-forming} = 12\pm2\,\text{m\AA} & 
\nonumber\\
A_{\rm quiescent} = 6\pm2\,\text{m\AA} & \mathrm{.}
\end{eqnarray}
In the right panel, we show the column density profiles normalized by the estimated virial radius of the galaxies. This removes the stellar mass dependence and the SFR dependence is slightly stronger with best-fit amplitudes:
\begin{eqnarray}
A'_{\rm star-forming} = 13\pm2\,\text{m\AA} & 
\nonumber\\
A'_{\rm quiescent} = 5\pm2\,\text{m\AA} & \mathrm{.}
\end{eqnarray}
This result indicates that either star formation (or a physical process correlated with it) is responsible for an excess of \caii\ in galaxy halos or that \caii\ is suppressed in the halo of quiescent galaxies.\\

\begin{figure}[!h]
\epsscale{1.2}
\plotone{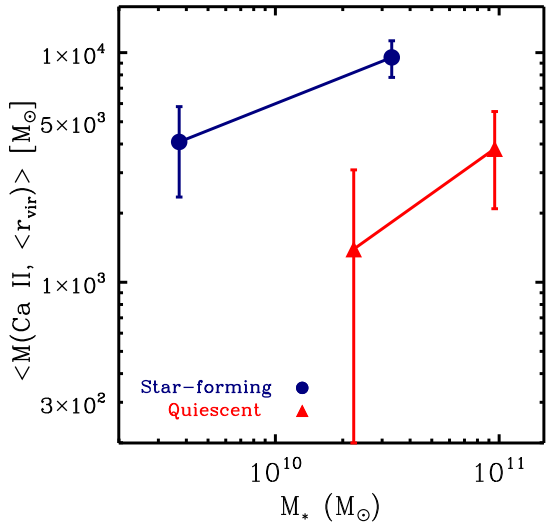}
\caption{Dependence of the total amount of \caii\ in the halo on stellar mass and SFR. The total amount of \caii\ is estimated by fitting the measured absorption radial profile with a power-law form with a fixed slope of $\alpha=-1.38$ and integrating the best-fit profile between $0.05\,r_{\rm vir}$ and $r_{\rm vir}$ ($10\,\kpc$ and $200\,\kpc$ for a $10^{10.3}\,\MSun$ galaxy).}
\label{fig:masssfrdepend}
\end{figure}

\begin{deluxetable}{cccc}
\tabletypesize{\scriptsize}
\tablecolumns{4}
\tablecaption{Average $\rm Ca\, II$ mass in the halo}
\tablehead{
 \colhead{Sample} &  \colhead{Median $M_*$} &  \colhead{Median $r_{\rm vir}$} & \colhead{$\langle M_{\rm Ca\,II}^{\rm Halo} \rangle$\tablenotemark{$a$}} \\
  \colhead{ } &  \colhead{[$10^{10}\,{\rm M_\odot}$]} & \colhead{[kpc]} & \colhead{[$10^{3}\,{\rm M_\odot}$]}  \\
}
\startdata
High-mass star-forming    & $3.3$   & $221$   & $9.5\pm1.8$  \\
Low-mass star-forming     & $0.4$  & $143$   & $4.1\pm1.8$  \\
High-mass quiescent       & $9.5$   & $274$   & $3.8\pm1.8$  \\
Low-mass quiescent        & $2.2$   & $205$   & $1.4\pm1.8$  \\
ALL    & $2.0$   & $200$   & $5.0\pm1.0$

\enddata
\tablenotetext{$a$}{Average total $\rm Ca\, II$ mass in the halo between $0.05\,r_{\rm vir}$ and $1\,r_{\rm vir}$ (between $10\,\kpc$ and $200\,\kpc$ for a $10^{10.3}\,M_\odot$ galaxy).}
\label{tbl:mass}
\end{deluxetable}


In Figure \ref{fig:masssfrdepend} and Table \ref{tbl:mass}, we summarize the total \caii\ mass in the halo (measured between $0.05\,r_{\rm vir}$ and $r_{\rm vir}$ using Eq.~\ref{eq:totalmass}) for the four galaxy subsamples. 
In short, we find that there is more \caii\ in the halo of galaxies with higher stellar mass and higher SFR. \\

$\bullet~$  {\bf Azimuthal Angle}: we now constrain the spatial distribution of \caii\ as a function of the location of the gas with respect to the disk of the galaxy. To do so we select edge-on galaxies with minor-to-major axis ratio $b/a<0.55$, and for a given galaxy-quasar pair we measure the azimuthal angle $\phi$ of the quasar from the galaxy's minor axis (determined from the position angle). We then divide the galaxy-quasar pairs into two subsamples: $|\phi|<45^\circ$ (along the minor axis) and $|\phi|>45^\circ$ (along the major axis) and measure the average \caii\ absorption for each subsample. Figure \ref{fig:azimuthal} shows the measured \caii\ absorption radial profiles. We find that, on average, the absorption is stronger along the minor axis (in the polar direction) than along the major axis (parallel to the disk plane). Fixing the slope $\alpha$, we find the best-fit amplitudes are:
\begin{eqnarray}
A_{\rm polar} = 18\pm3\,\text{m\AA} &
\nonumber\\
A_{\rm disk} = 6\pm3\,\text{m\AA} & \mathrm{.}
\end{eqnarray}
This indicates that there is more \caii\ perpendicular to the disk and indicate that bipolar outflows driven by star formation (or a process correlated with it) may be one of the major sources for \caii. We note that no significant excess is detected on large scales. It is likely that the azimuthal angle dependence itself depends on the galactocentric distance. \citet{bordoloi11a} reported a similar effect by measuring \mgii\ absorption around galaxies and was able to detect an azimuthal dependence only within $\sim50\,\kpc$ from the galaxies.

\section{Summary and discussion}

The \caii\ H \& K absorption lines have been used for two centuries to study a host of astrophysical phenomena.
In this paper we have extended the use of these lines to the circum- and inter-galactic contexts by measuring the galaxy-\caii\ gas spatial correlation function. Using spectra of about $10^5$ quasars from the SDSS DR7 dataset we have measured the mean \caii\ absorption around one million low-redshift galaxies with median $M_*\simeq10^{10.3}\,\MSun$ on scales ranging from about $10$ to $200\,\kpc$. This statistical technique allows us to probe absorption lines down to rest equivalent widths of several milli-Angstr\"oms and to be sensitive to the \emph{total} amount of \caii\ around galaxies.

The detection of absorption features relies on the ability to accurately estimate the intrinsic spectral energy distribution of the background source. To do so we have used the continuum estimates presented in Zhu \& M\'enard (2012), based on a vector decomposition technique. The analysis presented in this paper is based on the cross-correlation between galaxy positions and the flux decrements in the background quasar spectra. It has allowed us to obtain the following results:

\begin{figure}[t]
\epsscale{1.2}
\plotone{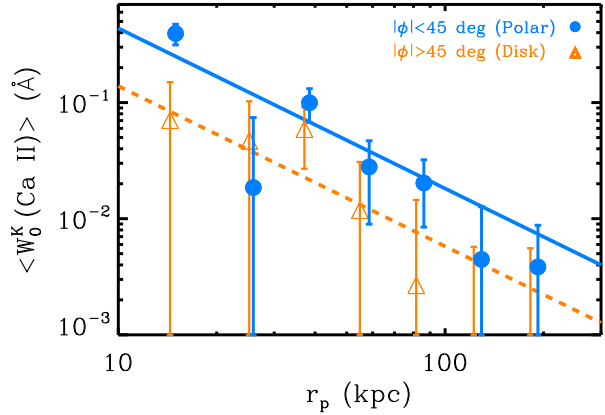}
\caption{Dependence of \caii\ absorption on azimuthal angle. The cyan solid circles show the \caii\ absorption along the minor axis, and the orange triangles along the major axis. There is more \caii\ along the minor axis. The corresponding lines show the best-fit power-law profiles with the slope fixed at $\alpha=-1.38$.}
\label{fig:azimuthal}
\end{figure}

\begin{itemize}

\item We detect \caii\ H \& K absorption out to projected galactocentric distance $r_{\mathrm p}\sim200~\kpc$ (about virial radius). This is about an order-of-magnitude farther out than previous studies based on individual absorber detections. Our sensitivity allows us to probe \caii\ column densities down to about $5\times10^{10}\,{\rm cm}^{-2}$.

\item The mean spatial distribution of \caii\ gas around galaxies follows a power-law column density profile with a slope $\alpha=-1.38\pm0.11$, at $r_\mathrm{p}<200\,\kpc$ (see Eq.~\ref{eq:profile}), which is steeper than an isothermal profile. We estimate the total \caii\ mass in the halo within $200\,\kpc$ to be $\sim5\times10^3\,\MSun$, averaged over the foreground galaxy population. This is an order-of-magnitude higher than the total \caii\ mass in the Milky Way disk, suggesting that more than $90\%$ of \caii\ in the Universe is in the circum- and inter-galactic environment.

\item The amount of \caii\ in the halo is larger for galaxies with higher stellar mass. We find that the mean \caii\ absorption at fixed virial radius-normalized impact parameter is independent of stellar mass. This suggests that the \caii\ mass in the halo is more closely related to the dark matter halo mass rather than the total stellar content of galaxies.

\item We find, at fixed stellar mass, more \caii\ in the halo of star-forming galaxies than quiescent ones by a factor of a few. 

\item For edge-on galaxies we find that \caii\ is preferentially distributed along the minor axis (in the polar direction). 

\end{itemize}

These results provide new constraints on the gas distribution around galaxies and bring new elements to our understanding of the cosmic baryon cycle. They also provide quantitative challenges for theoretical models for galaxy formation and evolution:

\begin{itemize}
\item[$\circ$] The SFR dependence of the amount of \caii\ is similar to that of the highly-ionized \ovi\ recently found by the HST COS-halos team \citep[][]{tumlinson11a}. These results imply star formation, or a process correlated with star formation, is responsible either for injecting \caii\ and \ovi\ into the halo or keeping a large amount of calcium and oxygen in the halo in certain physical condition. The azimuthal angle dependence of \caii\ and \mgii\ absorption (\citealt{bordoloi11a, kacprzak11b, bouche12a, bordoloi12a}, Rubin et al. in prep), which shows that low-ionization gas is more abundant and more kinematically violent in the polar direction of edge-on galaxies, suggests bipolar outflows driven by star formation must have played a significant role in injecting metals into the halo. This is also in agreement with the ubiquitous outflows found in post-starburst/star-forming galaxy spectra at various redshifts \citep[\eg][see \citealt{steidel10a} for more evidence and a review]{tremonti07a, weiner09a, rubin10b}. Following a period of star formation, the fate of \caii\ ions is unclear.

In contrast to the strong SFR dependence of metal lines, there appears to be no statistically significant difference between the \lymana\ absorption by neutral hydrogen around star-forming galaxies and that around quiescent ones \citep{thom12a}, suggesting neutral hydrogen and metals in the halo could have very different origins or be distributed separately in phase space.

\item[$\circ$] The amount of \caii\ in the halo is correlated with stellar mass. This 
dependence is shallower than linear, such that the total \caii\ mass is more closely related to dark matter mass than stellar mass. This may have the same cause as the mass-metallicity relation - the metallicity of interstellar gas decreases with stellar mass because metals are more easily expelled from low-mass galaxies due to their shallower gravitational potential. This will result in higher metal (halo)-to-stellar mass ratio for lower-mass galaxies. 
It is interesting to note that the total \ovi\ mass in the halo shows even less dependence on stellar mass \citep{tumlinson11a}.
\end{itemize}

Large and homogeneous surveys of the sky have allowed us to probe the distribution of matter in low-density environments. From the Sloan Digital Sky Survey only, we now have measurements of the galaxy-galaxy correlation function
from clustering analyses \citep[\eg][]{masjedi06a, jiang12a}, the galaxy-mass correlation from gravitational lensing \citep[\eg][]{mandelbaum06a}, the galaxy-dust correlation from reddening measurements \citep{menard10a} and, from the work presented in this paper, the galaxy-gas correlation function obtained by measuring statistical absorption by metals. These new observational probes are providing us with a more complete description of the matter distribution around galaxies.

This analysis shows the potential of using the ever-growing data from large surveys for absorption line studies. The automatic continuum estimation method presented in \citet{zhu13a} and the cross-correlation technique presented in this paper are generic and readily applicable to any large dataset from ongoing and future surveys such as BOSS \citep{schlegel09a}, eBOSS \citep[][]{comparat12a}, BigBOSS \citep[][]{schlegel11a}, and PFS \citep[][]{ellis12a}.

\acknowledgments

This work was supported by NSF Grant AST-1109665, the Alfred P. Sloan foundation and a grant from Theodore Dunham, Jr., Grant of Fund for Astrophysical Research. Funding for the SDSS and SDSS-II has been provided by the Alfred P. Sloan Foundation, the Participating Institutions, the National Science Foundation, the U.S. Department of Energy, the National Aeronautics and Space Administration, the Japanese Monbukagakusho, the Max Planck Society, and the Higher Education Funding Council for England. The SDSS Web Site is http://www.sdss.org/.

\bibliographystyle{apj}

\begin{thebibliography}{56}
\expandafter\ifx\csname natexlab\endcsname\relax\def\natexlab#1{#1}\fi

\bibitem[{{Abazajian} {et~al.}(2009){Abazajian}, {Adelman-McCarthy},
  {Ag{\"u}eros}, {Allam}, {Allende Prieto}, {An}, {Anderson}, {Anderson},
  {Annis}, {Bahcall}, \& et~al.}]{abazajian09a}
{Abazajian}, K.~N., {Adelman-McCarthy}, J.~K., {Ag{\"u}eros}, M.~A., {et~al.}
  2009, \apjs, 182, 543

\bibitem[{{Asplund} {et~al.}(2009){Asplund}, {Grevesse}, {Sauval}, \&
  {Scott}}]{asplund09a}
{Asplund}, M., {Grevesse}, N., {Sauval}, A.~J., \& {Scott}, P. 2009, \araa, 47,
  481

\bibitem[{{Behroozi} {et~al.}(2010){Behroozi}, {Conroy}, \&
  {Wechsler}}]{behroozi10a}
{Behroozi}, P.~S., {Conroy}, C., \& {Wechsler}, R.~H. 2010, \apj, 717, 379

\bibitem[{{Ben Bekhti} {et~al.}(2012){Ben Bekhti}, {Winkel}, {Richter}, {Kerp},
  {Klein}, \& {Murphy}}]{benbekhti12a}
{Ben Bekhti}, N., {Winkel}, B., {Richter}, P., {et~al.} 2012, \aap, 542, A110

\bibitem[{{Blades} {et~al.}(1981){Blades}, {Hunstead}, \&
  {Murdoch}}]{blades81a}
{Blades}, J.~C., {Hunstead}, R.~W., \& {Murdoch}, H.~S. 1981, \mnras, 194, 669

\bibitem[{{Blanton} \& {Roweis}(2007)}]{blanton07a}
{Blanton}, M.~R., \& {Roweis}, S. 2007, \aj, 133, 734

\bibitem[{{Boksenberg} {et~al.}(1980){Boksenberg}, {Danziger}, {Fosbury}, \&
  {Goss}}]{boksenberg80a}
{Boksenberg}, A., {Danziger}, I.~J., {Fosbury}, R.~A.~E., \& {Goss}, W.~M.
  1980, \apjl, 242, L145

\bibitem[{{Boksenberg} \& {Sargent}(1978)}]{boksenberg78a}
{Boksenberg}, A., \& {Sargent}, W.~L.~W. 1978, \apj, 220, 42

\bibitem[{{Bordoloi} {et~al.}(2012){Bordoloi}, {Lilly}, {Kacprzak}, \&
  {Churchill}}]{bordoloi12a}
{Bordoloi}, R., {Lilly}, S.~J., {Kacprzak}, G.~G., \& {Churchill}, C.~W. 2012,
  ArXiv e-prints

\bibitem[{{Bordoloi} {et~al.}(2011){Bordoloi}, {Lilly}, {Knobel}, {Bolzonella},
  {Kampczyk}, {Carollo}, {Iovino}, {Zucca}, {Contini}, {Kneib}, {Le Fevre},
  {Mainieri}, {Renzini}, {Scodeggio}, {Zamorani}, {Balestra}, {Bardelli},
  {Bongiorno}, {Caputi}, {Cucciati}, {de la Torre}, {de Ravel}, {Garilli},
  {Kova{\v c}}, {Lamareille}, {Le Borgne}, {Le Brun}, {Maier}, {Mignoli},
  {Pello}, {Peng}, {Perez Montero}, {Presotto}, {Scarlata}, {Silverman},
  {Tanaka}, {Tasca}, {Tresse}, {Vergani}, {Barnes}, {Cappi}, {Cimatti},
  {Coppa}, {Diener}, {Franzetti}, {Koekemoer}, {L{\'o}pez-Sanjuan},
  {McCracken}, {Moresco}, {Nair}, {Oesch}, {Pozzetti}, \&
  {Welikala}}]{bordoloi11a}
{Bordoloi}, R., {Lilly}, S.~J., {Knobel}, C., {et~al.} 2011, \apj, 743, 10

\bibitem[{{Bouch{\'e}} {et~al.}(2012){Bouch{\'e}}, {Hohensee}, {Vargas},
  {Kacprzak}, {Martin}, {Cooke}, \& {Churchill}}]{bouche12a}
{Bouch{\'e}}, N., {Hohensee}, W., {Vargas}, R., {et~al.} 2012, \mnras, 426, 801

\bibitem[{{Bowen} {et~al.}(1991){Bowen}, {Pettini}, {Penston}, \&
  {Blades}}]{bowen91a}
{Bowen}, D.~V., {Pettini}, M., {Penston}, M.~V., \& {Blades}, C. 1991, \mnras,
  249, 145

\bibitem[{{Brinchmann} {et~al.}(2004){Brinchmann}, {Charlot}, {White},
  {Tremonti}, {Kauffmann}, {Heckman}, \& {Brinkmann}}]{brinchmann04a}
{Brinchmann}, J., {Charlot}, S., {White}, S.~D.~M., {et~al.} 2004, \mnras, 351,
  1151

\bibitem[{{Bryan} \& {Norman}(1998)}]{bryan98a}
{Bryan}, G.~L., \& {Norman}, M.~L. 1998, \apj, 495, 80

\bibitem[{{Comparat} {et~al.}(2012){Comparat}, {Kneib}, {Escoffier}, {Zoubian},
  {Ealet}, {Lamareille}, {Mostek}, {Steele}, {Aubourg}, {Bailey}, {Bolton},
  {Brownstein}, {Dawson}, {Ge}, {Ilbert}, {Leauthaud}, {Maraston}, {Percival},
  {Ross}, {Schimd}, {Schlegel}, {Schneider}, {Thomas}, {Tinker}, \&
  {Weaver}}]{comparat12a}
{Comparat}, J., {Kneib}, J.-P., {Escoffier}, S., {et~al.} 2012, \mnras, 104

\bibitem[{{Draper}(1843)}]{draper43a}
{Draper}, W. 1843, The Philosophical Magazine and Journal of Science, 22

\bibitem[{{Ellis} {et~al.}(2012){Ellis}, {Takada}, {Aihara}, {Arimoto},
  {Bundy}, {Chiba}, {Cohen}, {Dore}, {Greene}, {Gunn}, {Heckman}, {Hirata},
  {Ho}, {Kneib}, {Le Fevre}, {Murayama}, {Nagao}, {Ouchi}, {Seiffert},
  {Silverman}, {Sodre}, {Spergel}, {Strauss}, {Sugai}, {Suto}, {Takami},
  {Wyse}, \& {the PFS Team}}]{ellis12a}
{Ellis}, R., {Takada}, M., {Aihara}, H., {et~al.} 2012, arXiv:1206.0737

\bibitem[{{Fraunhofer}(1814)}]{fraunhofer14a}
{Fraunhofer}, J. 1814, Denkschriften der K�niglichen Akademie der
  Wissenschaften zu M{\"u}nchen, 5, 193

\bibitem[{{Gudennavar} {et~al.}(2012){Gudennavar}, {Bubbly}, {Preethi}, \&
  {Murthy}}]{gudennavar12a}
{Gudennavar}, S.~B., {Bubbly}, S.~G., {Preethi}, K., \& {Murthy}, J. 2012,
  \apjs, 199, 8

\bibitem[{{Hartmann}(1904)}]{hartmann04a}
{Hartmann}, J. 1904, \apj, 19, 268

\bibitem[{{Hearnshaw}(1986)}]{hearnshaw86a}
{Hearnshaw}, J.~B. 1986, {The analysis of starlight: One hundred and fifty
  years of astronomical spectroscopy}

\bibitem[{{Hewett} \& {Wild}(2010)}]{hewett10a}
{Hewett}, P.~C., \& {Wild}, V. 2010, \mnras, 405, 2302

\bibitem[{{Jiang} {et~al.}(2012){Jiang}, {Hogg}, \& {Blanton}}]{jiang12a}
{Jiang}, T., {Hogg}, D.~W., \& {Blanton}, M.~R. 2012, \apj, 759, 140

\bibitem[{{Kacprzak} {et~al.}(2011){Kacprzak}, {Churchill}, {Evans}, {Murphy},
  \& {Steidel}}]{kacprzak11b}
{Kacprzak}, G.~G., {Churchill}, C.~W., {Evans}, J.~L., {Murphy}, M.~T., \&
  {Steidel}, C.~C. 2011, \mnras, 416, 3118

\bibitem[{{Kalberla} \& {Kerp}(2009)}]{kalberla09a}
{Kalberla}, P.~M.~W., \& {Kerp}, J. 2009, \araa, 47, 27

\bibitem[{{Kauffmann} {et~al.}(2003){Kauffmann}, {Heckman}, {Tremonti},
  {Brinchmann}, {Charlot}, {White}, {Ridgway}, {Brinkmann}, {Fukugita}, {Hall},
  {Ivezi{\'c}}, {Richards}, \& {Schneider}}]{kauffmann03a}
{Kauffmann}, G., {Heckman}, T.~M., {Tremonti}, C., {et~al.} 2003, \mnras, 346,
  1055

\bibitem[{{Lee} \& {Seung}(1999)}]{lee99a}
{Lee}, D.~D., \& {Seung}, H.~S. 1999, \nat, 401, 788

\bibitem[{{Lilly} {et~al.}(2007){Lilly}, {Le F{\`e}vre}, {Renzini}, {Zamorani},
  {Scodeggio}, {Contini}, {Carollo}, {Hasinger}, {Kneib}, {Iovino}, {Le Brun},
  {Maier}, {Mainieri}, {Mignoli}, {Silverman}, {Tasca}, {Bolzonella},
  {Bongiorno}, {Bottini}, {Capak}, {Caputi}, {Cimatti}, {Cucciati}, {Daddi},
  {Feldmann}, {Franzetti}, {Garilli}, {Guzzo}, {Ilbert}, {Kampczyk}, {Kovac},
  {Lamareille}, {Leauthaud}, {Borgne}, {McCracken}, {Marinoni}, {Pello},
  {Ricciardelli}, {Scarlata}, {Vergani}, {Sanders}, {Schinnerer}, {Scoville},
  {Taniguchi}, {Arnouts}, {Aussel}, {Bardelli}, {Brusa}, {Cappi}, {Ciliegi},
  {Finoguenov}, {Foucaud}, {Franceschini}, {Halliday}, {Impey}, {Knobel},
  {Koekemoer}, {Kurk}, {Maccagni}, {Maddox}, {Marano}, {Marconi}, {Meneux},
  {Mobasher}, {Moreau}, {Peacock}, {Porciani}, {Pozzetti}, {Scaramella},
  {Schiminovich}, {Shopbell}, {Smail}, {Thompson}, {Tresse}, {Vettolani},
  {Zanichelli}, \& {Zucca}}]{lilly07a}
{Lilly}, S.~J., {Le F{\`e}vre}, O., {Renzini}, A., {et~al.} 2007, \apjs, 172,
  70

\bibitem[{{Mandelbaum} {et~al.}(2006){Mandelbaum}, {Seljak}, {Kauffmann},
  {Hirata}, \& {Brinkmann}}]{mandelbaum06a}
{Mandelbaum}, R., {Seljak}, U., {Kauffmann}, G., {Hirata}, C.~M., \&
  {Brinkmann}, J. 2006, \mnras, 368, 715

\bibitem[{{Mascart}(1864)}]{mascart64a}
{Mascart}, {\'E}. 1864, in Annales Scientifiques de l'Ecole Normale Superieure,
  Vol.~1, 219--262

\bibitem[{{Masjedi} {et~al.}(2006){Masjedi}, {Hogg}, {Cool}, {Eisenstein},
  {Blanton}, {Zehavi}, {Berlind}, {Bell}, {Schneider}, {Warren}, \&
  {Brinkmann}}]{masjedi06a}
{Masjedi}, M., {Hogg}, D.~W., {Cool}, R.~J., {et~al.} 2006, \apj, 644, 54

\bibitem[{{M{\'e}nard} {et~al.}(2010){M{\'e}nard}, {Scranton}, {Fukugita}, \&
  {Richards}}]{menard10a}
{M{\'e}nard}, B., {Scranton}, R., {Fukugita}, M., \& {Richards}, G. 2010,
  \mnras, 405, 1025

\bibitem[{{Morton}(2003)}]{morton03a}
{Morton}, D.~C. 2003, \apjs, 149, 205

\bibitem[{{Moustakas} {et~al.}(2013){Moustakas}, {Coil}, {Aird}, {Blanton},
  {Cool}, {Eisenstein}, {Mendez}, {Wong}, {Zhu}, \& {Arnouts}}]{moustakas13a}
{Moustakas}, J., {Coil}, A., {Aird}, J., {et~al.} 2013, arXiv:1301.1688

\bibitem[{{Pasachoff} \& {Suer}(2010)}]{pasachoff10a}
{Pasachoff}, J.~M., \& {Suer}, T.-A. 2010, Journal of Astronomical History and
  Heritage, 13, 120

\bibitem[{{Putman} {et~al.}(2012){Putman}, {Peek}, \& {Joung}}]{putman12a}
{Putman}, M.~E., {Peek}, J.~E.~G., \& {Joung}, M.~R. 2012, \araa, 50, 491

\bibitem[{{Richter} {et~al.}(2011){Richter}, {Krause}, {Fechner}, {Charlton},
  \& {Murphy}}]{richter11a}
{Richter}, P., {Krause}, F., {Fechner}, C., {Charlton}, J.~C., \& {Murphy},
  M.~T. 2011, \aap, 528, A12

\bibitem[{{Routly} \& {Spitzer}(1952)}]{routly52a}
{Routly}, P.~M., \& {Spitzer}, Jr., L. 1952, \apj, 115, 227

\bibitem[{{Rubin} {et~al.}(2010){Rubin}, {Weiner}, {Koo}, {Martin},
  {Prochaska}, {Coil}, \& {Newman}}]{rubin10b}
{Rubin}, K.~H.~R., {Weiner}, B.~J., {Koo}, D.~C., {et~al.} 2010, \apj, 719,
  1503

\bibitem[{{Safronova} \& {Safronova}(2011)}]{safronova11a}
{Safronova}, M.~S., \& {Safronova}, U.~I. 2011, \pra, 83, 012503

\bibitem[{{Schlegel} {et~al.}(2009){Schlegel}, {White}, \&
  {Eisenstein}}]{schlegel09a}
{Schlegel}, D., {White}, M., \& {Eisenstein}, D. 2009, in ArXiv Astrophysics
  e-prints, Vol. 2010, astro2010: The Astronomy and Astrophysics Decadal
  Survey, 314

\bibitem[{{Schlegel} {et~al.}(2011){Schlegel}, {Abdalla}, {Abraham}, {Ahn},
  {Allende Prieto}, {Annis}, {Aubourg}, {Azzaro}, {Baltay}, {Baugh}, {Bebek},
  {Becerril}, {Blanton}, {Bolton}, {Bromley}, {Cahn}, {Carton},
  {Cervantes-Cota}, {Chu}, {Cortes}, {Dawson}, {Dey}, {Dickinson}, {Diehl},
  {Doel}, {Ealet}, {Edelstein}, {Eppelle}, {Escoffier}, {Evrard}, {Faccioli},
  {Frenk}, {Geha}, {Gerdes}, {Gondolo}, {Gonzalez-Arroyo}, {Grossan},
  {Heckman}, {Heetderks}, {Ho}, {Honscheid}, {Huterer}, {Ilbert}, {Ivans},
  {Jelinsky}, {Jing}, {Joyce}, {Kennedy}, {Kent}, {Kieda}, {Kim}, {Kim},
  {Kneib}, {Kong}, {Kosowsky}, {Krishnan}, {Lahav}, {Lampton}, {LeBohec}, {Le
  Brun}, {Levi}, {Li}, {Liang}, {Lim}, {Lin}, {Linder}, {Lorenzon}, {de la
  Macorra}, {Magneville}, {Malina}, {Marinoni}, {Martinez}, {Majewski},
  {Matheson}, {McCloskey}, {McDonald}, {McKay}, {McMahon}, {Menard},
  {Miralda-Escude}, {Modjaz}, {Montero-Dorta}, {Morales}, {Mostek}, {Newman},
  {Nichol}, {Nugent}, {Olsen}, {Padmanabhan}, {Palanque-Delabrouille}, {Park},
  {Peacock}, {Percival}, {Perlmutter}, {Peroux}, {Petitjean}, {Prada},
  {Prieto}, {Prochaska}, {Reil}, {Rockosi}, {Roe}, {Rollinde}, {Roodman},
  {Ross}, {Rudnick}, {Ruhlmann-Kleider}, {Sanchez}, {Sawyer}, {Schimd},
  {Schubnell}, {Scoccimaro}, {Seljak}, {Seo}, {Sheldon}, {Sholl},
  {Shulte-Ladbeck}, {Slosar}, {Smith}, {Smoot}, {Springer}, {Stril}, {Szalay},
  {Tao}, {Tarle}, {Taylor}, {Tilquin}, {Tinker}, {Valdes}, {Wang}, {Wang},
  {Weaver}, {Weinberg}, {White}, {Wood-Vasey}, {Yang}, {Yeche}, {Zakamska},
  {Zentner}, {Zhai}, \& {Zhang}}]{schlegel11a}
{Schlegel}, D., {Abdalla}, F., {Abraham}, T., {et~al.} 2011, arXiv:1106.1706


\bibitem[{{Schneider} {et~al.}(1993){Schneider}, {Hartig}, {Jannuzi},
  {Kirhakos}, {Saxe}, {Weymann}, {Bahcall}, {Bergeron}, {Boksenberg},
  {Sargent}, {Savage}, {Turnshek}, \& {Wolfe}}]{schneider93a}
{Schneider}, D.~P., {Hartig}, G.~F., {Jannuzi}, B.~T., {et~al.} 1993, \apjs,
  87, 45

\bibitem[{{Schneider} {et~al.}(2010){Schneider}, {Richards}, {Hall}, {Strauss},
  {Anderson}, {Boroson}, {Ross}, {Shen}, {Brandt}, {Fan}, {Inada}, {Jester},
  {Knapp}, {Krawczyk}, {Thakar}, {Vanden Berk}, {Voges}, {Yanny}, {York},
  {Bahcall}, {Bizyaev}, {Blanton}, {Brewington}, {Brinkmann}, {Eisenstein},
  {Frieman}, {Fukugita}, {Gray}, {Gunn}, {Hibon}, {Ivezi{\'c}}, {Kent}, {Kron},
  {Lee}, {Lupton}, {Malanushenko}, {Malanushenko}, {Oravetz}, {Pan}, {Pier},
  {Price}, {Saxe}, {Schlegel}, {Simmons}, {Snedden}, {SubbaRao}, {Szalay}, \&
  {Weinberg}}]{schneider10a}
{Schneider}, D.~P., {Richards}, G.~T., {Hall}, P.~B., {et~al.} 2010, \aj, 139,
  2360

\bibitem[{{Steidel} {et~al.}(2010){Steidel}, {Erb}, {Shapley}, {Pettini},
  {Reddy}, {Bogosavljevi{\'c}}, {Rudie}, \& {Rakic}}]{steidel10a}
{Steidel}, C.~C., {Erb}, D.~K., {Shapley}, A.~E., {et~al.} 2010, \apj, 717, 289

\bibitem[{{Thilker} {et~al.}(2004){Thilker}, {Braun}, {Walterbos}, {Corbelli},
  {Lockman}, {Murphy}, \& {Maddalena}}]{thilker04a}
{Thilker}, D.~A., {Braun}, R., {Walterbos}, R.~A.~M., {et~al.} 2004, \apjl,
  601, L39

\bibitem[{{Thom} {et~al.}(2012){Thom}, {Tumlinson}, {Werk}, {Prochaska},
  {Oppenheimer}, {Peeples}, {Tripp}, {Katz}, {O'Meara}, {Brady Ford},
  {Dav{\'e}}, {Sembach}, \& {Weinberg}}]{thom12a}
{Thom}, C., {Tumlinson}, J., {Werk}, J.~K., {et~al.} 2012, \apjl, 758, L41

\bibitem[{{Tremonti} {et~al.}(2007){Tremonti}, {Moustakas}, \&
  {Diamond-Stanic}}]{tremonti07a}
{Tremonti}, C.~A., {Moustakas}, J., \& {Diamond-Stanic}, A.~M. 2007, \apjl,
  663, L77

\bibitem[{{Tumlinson} {et~al.}(2011){Tumlinson}, {Thom}, {Werk}, {Prochaska},
  {Tripp}, {Weinberg}, {Peeples}, {O'Meara}, {Oppenheimer}, {Meiring}, {Katz},
  {Dav{\'e}}, {Ford}, \& {Sembach}}]{tumlinson11a}
{Tumlinson}, J., {Thom}, C., {Werk}, J.~K., {et~al.} 2011, Science, 334, 948

\bibitem[{{Wakker}(2001)}]{wakker01a}
{Wakker}, B.~P. 2001, \apjs, 136, 463

\bibitem[{{Wakker} \& {Mathis}(2000)}]{wakker00a}
{Wakker}, B.~P., \& {Mathis}, J.~S. 2000, \apjl, 544, L107

\bibitem[{{Weiner} {et~al.}(2009){Weiner}, {Coil}, {Prochaska}, {Newman},
  {Cooper}, {Bundy}, {Conselice}, {Dutton}, {Faber}, {Koo}, {Lotz}, {Rieke}, \&
  {Rubin}}]{weiner09a}
{Weiner}, B.~J., {Coil}, A.~L., {Prochaska}, J.~X., {et~al.} 2009, \apj, 692,
  187

\bibitem[{{Welty} {et~al.}(1996){Welty}, {Morton}, \& {Hobbs}}]{welty96a}
{Welty}, D.~E., {Morton}, D.~C., \& {Hobbs}, L.~M. 1996, \apjs, 106, 533

\bibitem[{{York} {et~al.}(2000){York}, {Adelman}, {Anderson}, {Anderson},
  {Annis}, {Bahcall}, {Bakken}, {Barkhouser}, {Bastian}, {Berman}, {Boroski},
  {Bracker}, {Briegel}, {Briggs}, {Brinkmann}, {Brunner}, {Burles}, {Carey},
  {Carr}, {Castander}, {Chen}, {Colestock}, {Connolly}, {Crocker}, {Csabai},
  {Czarapata}, {Davis}, {Doi}, {Dombeck}, {Eisenstein}, {Ellman}, {Elms},
  {Evans}, {Fan}, {Federwitz}, {Fiscelli}, {Friedman}, {Frieman}, {Fukugita},
  {Gillespie}, {Gunn}, {Gurbani}, {de Haas}, {Haldeman}, {Harris}, {Hayes},
  {Heckman}, {Hennessy}, {Hindsley}, {Holm}, {Holmgren}, {Huang}, {Hull},
  {Husby}, {Ichikawa}, {Ichikawa}, {Ivezi{\'c}}, {Kent}, {Kim}, {Kinney},
  {Klaene}, {Kleinman}, {Kleinman}, {Knapp}, {Korienek}, {Kron}, {Kunszt},
  {Lamb}, {Lee}, {Leger}, {Limmongkol}, {Lindenmeyer}, {Long}, {Loomis},
  {Loveday}, {Lucinio}, {Lupton}, {MacKinnon}, {Mannery}, {Mantsch}, {Margon},
  {McGehee}, {McKay}, {Meiksin}, {Merelli}, {Monet}, {Munn}, {Narayanan},
  {Nash}, {Neilsen}, {Neswold}, {Newberg}, {Nichol}, {Nicinski}, {Nonino},
  {Okada}, {Okamura}, {Ostriker}, {Owen}, {Pauls}, {Peoples}, {Peterson},
  {Petravick}, {Pier}, {Pope}, {Pordes}, {Prosapio}, {Rechenmacher}, {Quinn},
  {Richards}, {Richmond}, {Rivetta}, {Rockosi}, {Ruthmansdorfer}, {Sandford},
  {Schlegel}, {Schneider}, {Sekiguchi}, {Sergey}, {Shimasaku}, {Siegmund},
  {Smee}, {Smith}, {Snedden}, {Stone}, {Stoughton}, {Strauss}, {Stubbs},
  {SubbaRao}, {Szalay}, {Szapudi}, {Szokoly}, {Thakar}, {Tremonti}, {Tucker},
  {Uomoto}, {Vanden Berk}, {Vogeley}, {Waddell}, {Wang}, {Watanabe},
  {Weinberg}, {Yanny}, \& {Yasuda}}]{york00a}
{York}, D.~G., {Adelman}, J., {Anderson}, Jr., J.~E., {et~al.} 2000, \aj, 120,
  1579

\bibitem[{{York} {et~al.}(2012){York}, {Straka}, {Bishof}, {Kuttruff}, {Bowen},
  {Kulkarni}, {Subbarao}, {Richards}, {Vanden Berk}, {Hall}, {Heckman},
  {Khare}, {Quashnock}, {Ghering}, \& {Johnson}}]{york12a}
{York}, D.~G., {Straka}, L.~A., {Bishof}, M., {et~al.} 2012, \mnras, 423, 3692

\bibitem[{{Zhu} \& {M{\'e}nard}(2012)}]{zhu13a}
{Zhu}, G., \& {M{\'e}nard}, B. 2012, arXiv:1211.6215


\end{thebibliography}

\end{document}